\documentclass{aa}  
\usepackage{graphicx}

\usepackage{txfonts}
\begin{document}
   \title{Direct detection of a magnetic field in the photosphere of the single M giant EK~Boo\thanks{Based on data obtained using the T\'elescope Bernard Lyot at Observatoire du Pic du Midi, CNRS and Universit\'e de Toulouse, France.}}
  \subtitle{How common is magnetic activity among M giants?}

   \author{R. Konstantinova-Antova\inst{1,2}, M. Auri\`ere\inst{2}, C. Charbonnel\inst{3,2}, 
N.A. Drake\inst{4}, 
K.-P. Schr\"oder\inst{5}, I. Stateva\inst{1}, E. Alecian\inst{6}, P. Petit\inst{2}, and R.Cabanac\inst{2}}

   \offprints{R. Konstantinova-Antova}

   \institute{Institute of Astronomy, Bulgarian Academy of Sciences, 72 Tsarigradsko shose, 1784 Sofia,
             Bulgaria\\
              \email{renada@astro.bas.bg}              
\and
 Laboratoire d'Astrophysique de Toulouse-Tarbes, Universit\'e de Toulouse, CNRS, Observatoire Midi Pyr\'en\'es,  57 
Avenue d'Azereix, 65008 Tarbes, France
\and
 Observatoire Astronomique de l'Universit\'e de Gen\`eve, 51, Chemin des Maillettes, 1290 Versoix, Switzerland
\and
Sobolev Astronomical Institute, St. Petersburg State University, Universitetski pr. 28, St. Petersburg 198504, Russia
 \and
   Departamento de Astronomia, Universidad de Guanajuato, A.P. 144, C.P. 36000, GTO, Mexico
\and
Laboratoire d'Astrophysique de Grenoble, 
Universit\'e Joseph Fourier-CNRS, BP 53, 38041 Grenoble Cedex 9, France 
}

   \date{Received ; accepted }

\abstract 
{}
{We study the fast rotating M5 giant EK~Boo by means of  
spectropolarimetry to obtain direct and simultaneous 
measurements of both the magnetic field and activity indicators, in 
order to infer the origin of  the activity in this fairly evolved giant.}
{We used the new spectropolarimeter NARVAL  at the Bernard Lyot Telescope
(Observatoire du Pic du Midi, France) to obtain a series of Stokes $I$ and 
Stokes $V$ profiles for EK~Boo. Using the Least Square Deconvolution technique we were able to 
detect the Zeeman signature of the magnetic field. We measured its 
longitudinal component by means of the averaged Stokes V and Stokes I 
profiles. The spectra also permitted us to monitor the Ca\,{\sc ii} 
K\&H chromospheric emission lines, which are well known as indicators
of stellar magnetic activity.}
{From ten observations obtained between April 2008 and March 2009, 
	we deduce that EK~Boo has a magnetic field, which varied in the 
range of -0.1 to -8 G. On March 13, 2009, a complex structure of Stokes $V$ 
was observed, which might indicate a dynamo. We also determined the initial 
mass and evolutionary stage of EK~Boo, based on up-to-date stellar evolution 
tracks. The initial mass is in the range of 2.0-3.6~M$_{\odot}$, and EK~Boo
is either on the asymptotic giant branch (AGB), at the onset of the thermal 
pulse phase, or at the tip of the first (or red) giant branch (RGB). The 
fast rotation and activity of EK~Boo might be explained by angular 
momentum dredge--up from the interior, or by the merging of a binary. 

In addition, we observed eight other M giants, which are known as X-ray emitters,
or to be rotating fast for their class. For one of these, $\beta$~And, 
presumably also an AGB star, we have a marginal detection of magnetic field, 
and a longitudinal component $B_l$ of about 1G was measured. More observations like this
will answer the question whether EK~Boo is a special case, 
or whether magnetic activity is, rather, more common among M giants
than expected.}
{}{}

   \authorrunning {Konstantinova-Antova et al.}
   \titlerunning {Direct detection of magnetic field in the single M giant EK~Boo}
   \maketitle 


\section{Introduction}

There are already measurements of magnetic fields in single G and K 
giants (Konstantinova--Antova et al.2008; Auri\`ere et al.2008; 
Konstantinova--Antova et al.2009), but M giants have not yet been studied in 
this respect. Despite theoretical predictions of possible dynamo 
operation on the asymptotic giant branch (AGB), (Blackman et al. 2001; 
Soker \& Zoabi 2002; Nordhaus et al. 2008), actual evidence of magnetic 
activity in such evolved stars is sparse and indirect (H\"unsch et al. 1998; 
Karovska et al. 2005; Herpin et al. 2006). Hence, we here present a study of 
the variable magnetic field of the single M giant EK~Boo. This work is a part 
of a programme, that aims to detect magnetic fields in evolved, 
single late-type stars (Konstantinova-Antova et al. 2008).

The source EK~Boo = HD~130144 is a 6th magnitude M5 giant and a semiregular variable star 
(Kholopov et al. 1998). This evolved giant attracted our interest by its 
unusual X-ray luminosity $L_x$, which is  higher than $10^{30}$ erg$s^{-1}$ 
(H\"unsch et al. 1998). A further study revealed its variability of $L_x$, 
as well as its optical activity indicators Ca\,{\sc ii} K\&H and 
H$_\alpha$ (H\"unsch et al. 2004). At the same time, EK~Boo is considered 
to be a single star (Famaey et al. 2009) with a projected rotation velocity  
of $v\sin i=11$~km\,s$^{-1}$ (H\"unsch et al. 2004). A direct detection of 
its magnetic field was first reported by Konstantinova-Antova 
et al. (2009). To our knowledge, this was the first direct detection 
of a magnetic field in a single M giant. In the work we present here, 
we study the variability of the magnetic field and activity of EK~Boo, 
its chemical composition and evolutionary stage, and we discuss possible 
reasons for a dynamo operation in such an evolved star. To see if this magnetically active M giant is a special case, or magnetic activity is common in such stars, we also present first magnetic field 
measurements obtained for other M giants.

\section{Observations and methods}

The observations of EK~Boo were carried out at the 2-m Bernard Lyot 
Telescope (TBL) of the Pic du Midi observatory, with NARVAL, a new 
generation spectropolarimeter (Auri\`ere \cite{auriere03}). It is 
a copy of the instrument ESPaDOnS, which was installed at CFHT at the 
end of 2004 (Donati et al. \cite{donati06}). NARVAL is a fiber-fed 
echelle spectrometer, able to cover the whole spectrum from 370 nm to 1000 nm 
in a single exposure. Forty orders are aligned on the CCD frame, 
separated by 2 cross-disperser prisms. We used NARVAL in polarimetric 
mode, with a spectral resolution of 65\,000. Stokes $I$ (unpolarised) and 
Stokes $V$ (circular polarization) parameters were measured by means of
four sub-exposures, between which the retarders, Fresnel rhombs, were rotated 
in order to exchange the beams in the instrument and to reduce spurious 
polarization signatures.

We observed EK~Boo during four consecutive nights in April 2008, then on 
20 and 21 December 2008, 25 February 2009, and during three nights in March 2009. 
For the extraction of the spectra we used Libre-ESpRIT (Donati et al. 
\cite{donati97}), a fully automatic reduction package installed at 
TBL. For the Zeeman analysis, a least-squares deconvolution 
(LSD, Donati et al. \cite{donati97}) was then applied to all spectra. 
We used a mask, which was calculated for solar abundance, an effective 
temperature of 3500~K, $\log g =0.5$ and a microturbulence of 2.0 $km s^{-1}$. 
These parameters are consistent with the spectral class and luminosity of 
EK~Boo (Dyck et al. 1998). This method enabled us to average a total of about 
12,700 lines and to get Stokes $I$ and Stokes $V$ profiles. 
The null spectrum ($N$ profile) given by the standard procedure 
(Donati et al. \cite{donati97}) was also examined, but never showed any 
signal, confirming that the detected signatures in the V profiles are not spurious. From the obtained mean Stokes $V$ for each night we  computed the 
surface-averaged longitudinal magnetic field $B_{l}$ in G, using the 
first-order moment method (Donati et al. \cite{donati97}, Rees \& Semel 
\cite{rees79}). 

At the same time, the magnetic activity of EK Boo was monitored by means 
of measurements of the relative intensity of the chromospheric activity 
indicator Ca\,{\sc ii}~K. In the core of this emission line, 
the $S/N$ of our spectra exceeds 50. Table 1 gives the journal of 
our observations of EK~Boo, including the dates, Heliocentric Julian Day (HJD), total exposure time, 
detection-level from the LSD statistics (Donati et al. \cite{donati97}), 
$B_{l}$ and its error (in G), and the relative intensity I(CaII K)/I(3950) 
for the maximum of the Ca\,{\sc ii} K emission.

\section{Observational results}
\subsection{Magnetic field and activity indicators of EK~Boo}

 A significant Stokes $V$ signal was detected at all times, except in December 
2008. The measured $B_{l}$ varies from $-0.1$ to --8~G. Figure~1 shows our 
best LSD profiles, obtained on 13 March 2009, under excellent sky conditions. 
Using the LSD statistics tool (Donati et al. 1997), the outstanding Stokes 
$V$ signature (upper profile) corresponds to a definitive detection (DD) 
with a $\chi^2$ of 4.7. The Stokes $V$ signal is well centred on the 
Stokes $I$ profile (bottom) and no signal at all is observed on the null 
polarization $N$ profile (middle). This observation shows a complex 
Stokes $V$ structure, maybe a result of a dynamo action. Figure 2 shows all 
observed Stokes $V$ profiles for EK~Boo. A very weak signal was observed 
in the two nights of December 2008, and we did not obtain any significant 
Zeeman detection. Nevertheless, a definitive detection can be attained by 
averaging all spectra taken during these two nights. On 25 February 
2009, the magnetic field was definitely detected again and the resulting 
$B_{l}$ is still close to 0. This challenges our much higher measurements 
obtained in the four nights of April 2008, when the magnetic field of EK~Boo 
did not present significant variations. And there appears a rapid increase 
of $B_l$ again in March 2009.

A significant Ca\,{\sc ii} K\&H emission is observed in every night. The 
ratio I/I(3950) of the K-line is given in Table~1. In Fig.~3 it can be 
compared with the $B_l$ measurements. Both measurements are plotted over 
HJD. We can see that the Ca\,{\sc ii} K emission 
increases smoothly and monotonically during the period monitored here, 
apparently without any significant short-time variation. This behaviour 
is different from that of the Stokes $V$ profiles and $B_l$. For example, 
Ca\,{\sc ii} is at the weakest when the magnetic field is strong in the
nights in April 2008. And it becomes stronger when Stokes V 
and $B_l$ are weak on 20 and 21 December 2008. This suggest the existence 
of areas with opposite magnetic polarity on the surface of EK~Boo:  
Contributions of both polarities to the mean magnetic field can cancel 
each other when both are present on the facing side of the star, while 
adding up to a maximum Ca\,{\sc ii} emission line flux of the related
chromospheric activity regions.

\begin{table}
\caption{Journal of observations of EK~Boo }             
\label{table:1}      
\centering                          
\begin{tabular}{c c c c c c c}        
\hline\hline                 
Date     &HJD      & Exp. T.   &Det.     & B$_l$& $\sigma$ &Ca\,{\sc ii} K \\
         &2450000+ & s         &         & G    & G        &       \\
\hline                        
03 Apr 08& 4560.50 & 800       &DD       &-6.7  &1.8       & 0.39  \\
04 Apr 08& 4561.53 & 1600      &DD       &-3.1  &0.5       & 0.39  \\
05 Apr 08& 4562.54 & 1600      &DD       &-5.0  &0.5       & 0.40  \\
06 Apr 08& 4563.51 & 1600      &DD       &-4.7  &0.7       & 0.36  \\
20 Dec 08& 4821.74 & 2400      &nd       &-0.1  &0.6       & 0.46  \\
21 Dec 08& 4822.74 & 2800      &nd       &-0.5  &0.6       & 0.49  \\
25 Feb 09& 4888.6  & 3200      &DD       &-0.3  &0.4       & 0.57  \\
09 Mar 09& 4900.57 & 3200      &DD       &-3.8  &0.6       & 0.55  \\
13 Mar 09& 4904.68 & 3200      &DD       &-4.6  &0.4       & 0.56  \\
18 Mar 09& 4909.55 & 3200      &DD       &-8.1  &0.6       & 0.57  \\
\hline  
         &         &           &         &      &          &       \\
\end{tabular}
Individual columns correspond to Dates, HJD, total exposure time, Zeeman 
detection (DD = definitive detection, nd = no detection), $B_l$ values 
and its errors in G, and I(CaII K)/I(3950) for Ca\,{\sc ii} K emission.
\end{table}

   \begin{figure}
   \centering
   \includegraphics[width=7cm, angle=0]{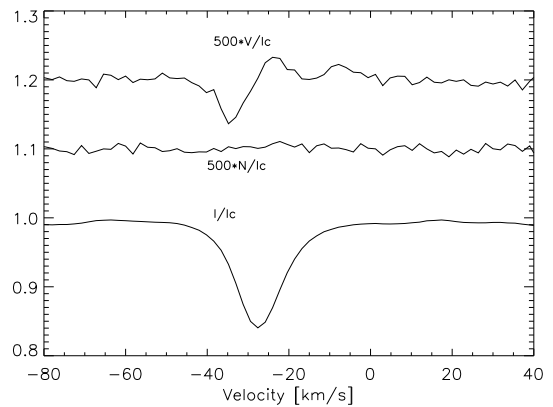}
      \caption{LSD profiles of Stokes $I$, null polarization $N$ and 
Stokes $V$ (bottom, middle and upper curves) for EK~Boo, as observed 
on 13 March 2009. For clarity, Stokes $V$ and $N$ profiles are 
enlarged 500 times, and the successive profiles are shifted vertically.}
         \label{Figure1}
   \end{figure}

   \begin{figure}
   \centering
   \includegraphics[width=7.4cm, angle=0]{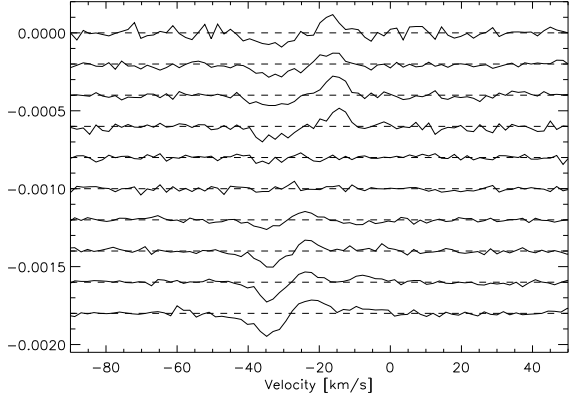}
     \caption{LSD Stokes $V$ profiles for EK~Boo for the dates listed
in Table~1, from 03 Apr 08 (top) to 18 Mar 09 (bottom). The dashed lines 
illustrate the respective zero levels. The scale of the Y-axis is in units 
of $V/I_c$, $I_c$ is the intensity of the continuum. For clarity, successive 
profiles are shifted vertically.}
         \label{Figure2}
   \end{figure}
   
The Ca\,{\sc ii} K emission profiles sometimes feature a single peak, 
but on other occations they show a double-peak structure (see Fig.~4). 
Their peak asymmetry changes from $V/R > 1$ on April 3, 2008, to $V/R < 1$ 
on December 21, 2008, where V/R is the ratio between the violet (V) and red (R) components
of the double-peaked emission line. In the first case, this indicates vertical motions 
in the chromosphere, where the downward components appear stronger. This 
is typical for many red giants (Smith \& Shetrone 2000). In the second 
case, $V/R < 1$, upward motions and possible outflow dominate the asymmetry. 
Hence, the chromosphere of EK~Boo seems to be very dynamical on timescales 
of months and years, with episodic outflows from its lower layers, 
where Ca\,{\sc ii} K and H emission lines are formed.

  \begin{figure}
   \centering
   \includegraphics[width=7cm, height=7.8cm, angle=270]{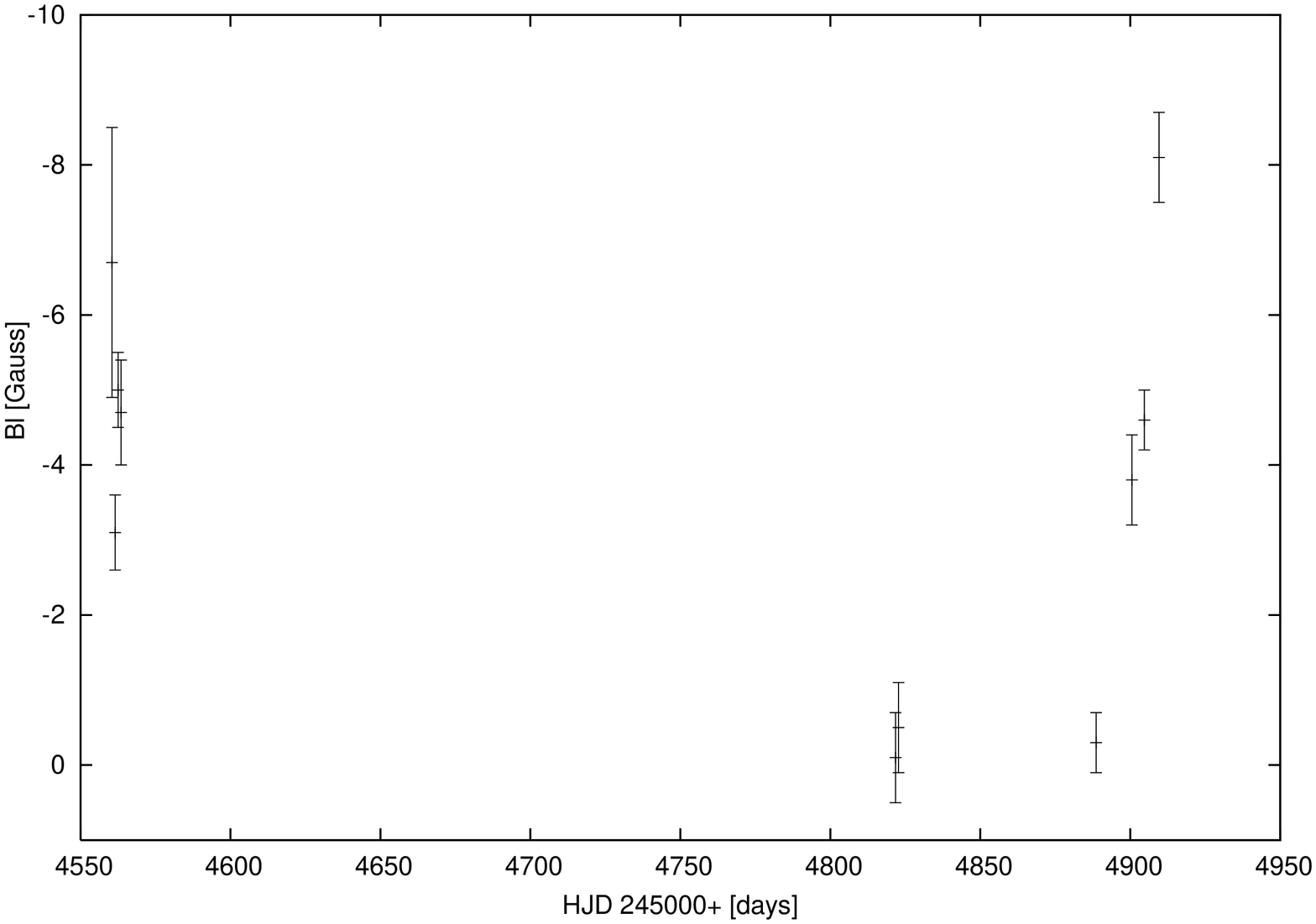}
   \includegraphics[width=7cm, height=8cm, angle=270]{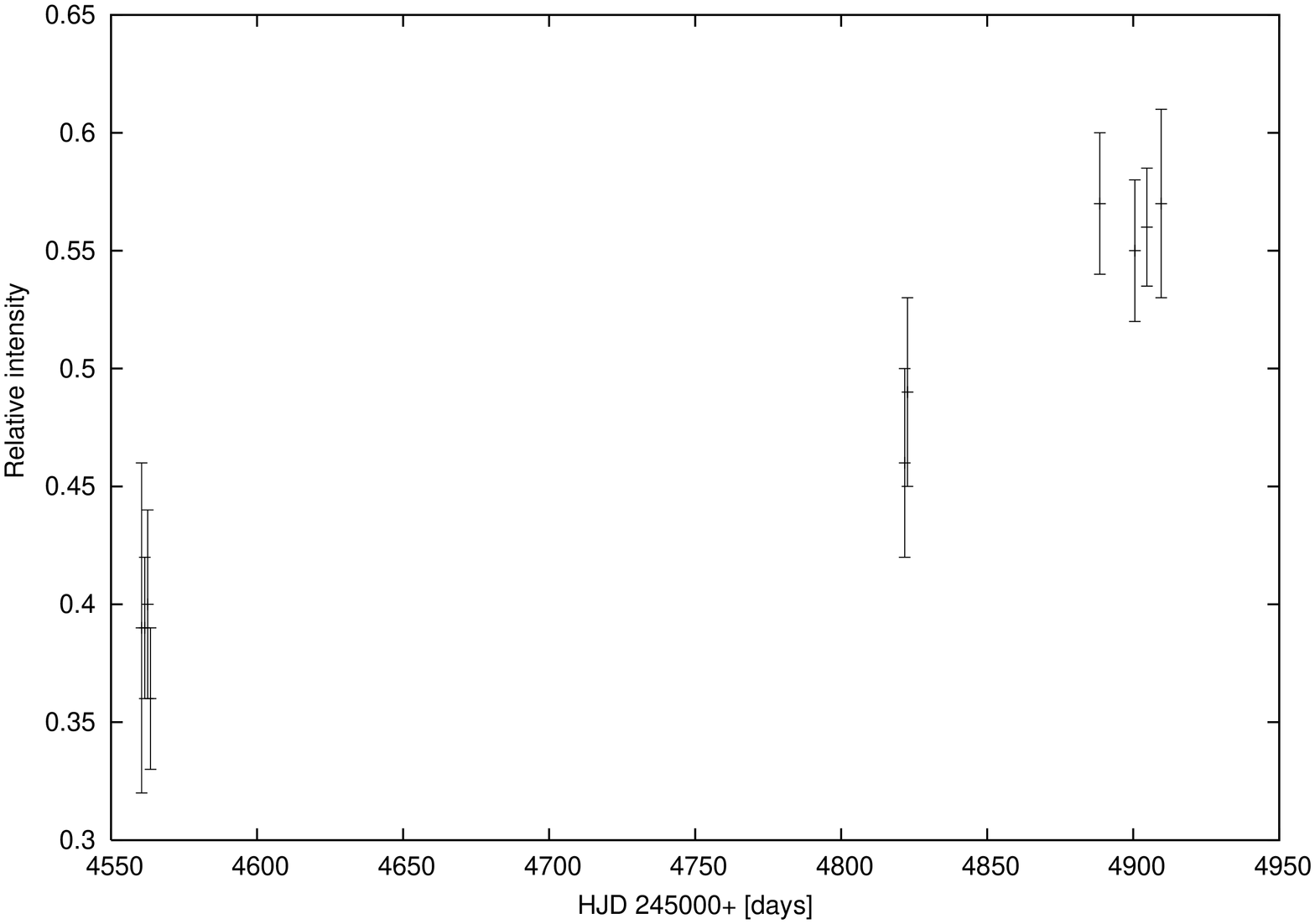}
      \caption{Magnetic field measurements (upper panel) and chromospheric 
activity indicator Ca\,{\sc ii} K emission (lower panel) 
for EK~Boo in the observed period. An inverse Y-axis is used for 
B$_l$ to illustrate the magnetic field. }
         \label{Figure3}
   \end{figure}
  \begin{figure}
   \centering
   \includegraphics[width=6.3cm, angle=270]{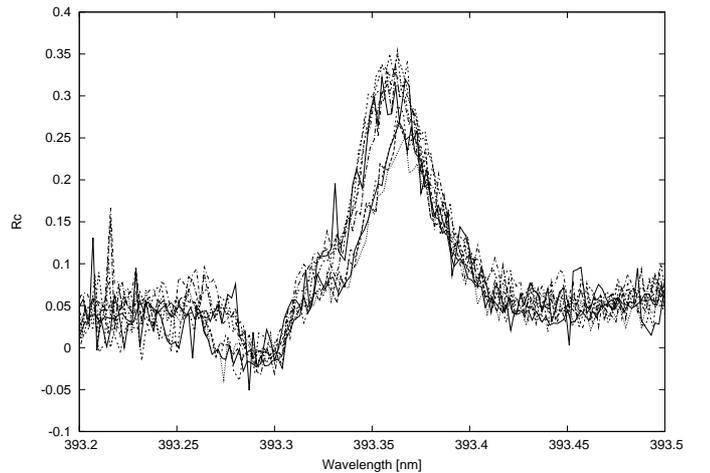}
   \caption{Ca\,{\sc ii} K emission line profiles, as observed for 
EK~Boo in the period of April 2008 to March 2009.}
         \label{Figure4}
   \end{figure}

\subsection{Radial velocity, binarity, and pulsations}

 Radial velocity measurements ($RV$) for EK~Boo have been carried out on 
the LSD Stokes $I$ profiles using Gaussian fits. The $RV$ accuracy of 
NARVAL is 20-30~m\,s$^{-1}$ (Moutou et al. 2007, Auri\`ere et al. 2009). 
Variations of $RV$ larger than 1~km\,s$^{-1}$ were observed on a time-scale 
of less than a month. But in averaging the $RV$ measurements for each month 
(April 2008, December 2008, February 2009, and March 2009), we find a 
standard deviation of only about half a km\,s$^{-1}$. This compares well 
with the values (0.36 and 0.65 km\,s$^{-1}$ respectively) obtained by 
H\"unsch et al. (2004) and Famaey et al. (2009) from measurements with 
CORAVEL (Udry et al. 1997). Famaey et al. (2009) included their EK~Boo 
observations in two of their diagnosis plots and concluded that the star 
``appears amidst the non-binary M giants''. 

EK~Boo is also known as a semi-regular variable (Kholopov et al. 1998). 
A larger photometric sample is available from Hipparcos data, and though 
EK~Boo is among stars with large HIP photometric amplitudes (0.38 mag., 
Adelman 2001), no clear periodicity can be derived (Koen et al. 2002). 

The average of our much more recent $RV$ measurement is shifted by 
about 5~km\,s$^{-1}$ with respect to the average given in the literature 
(about --27~km\,s$^{-1}$ with respect to about --22~km\,s$^{-1}$). This 
difference is one order of magnitude larger than any shift obtained for 
other M giants studied with NARVAL, even when observations were made during 
the same nights as EK~Boo. Hence, this shift appears to be real, and it 
could be linked to pulsations, as discussed in the context of other 
semiregular variable stars (Lebzelter 1999, Lebzelter et al. 2000) or 
of multiple-mode asymptotic giant branch variable stars (Hinkle et al. 2002). 

In summary, our $RV$ measurements confirm that their variations on a time-scale of less than a month are of a magnitude of about 1~km\,s$^{-1}$. These variations 
must be considered as significant with respect to NARVAL's accuracy (better 
than CORAVEL accuracy). The analysis by Famaey et al. (2009) favours an origin 
from pulsations, quite common in this type of stars. Furthermore, the shift in 
$RV$ between our observations and those in the literature is not due to 
binarity, although slow orbital motion might, a priory, be a possible 
explanation: H\"unsch et al. (1998) already considered the possibility of 
active, faint companions in detail, when discussing the origin of detected 
X-ray emission in M-giants. But in the case of EK~Boo, spectral signatures 
of a secondary have never been found. On the other hand, clearly  
a Zeeman detection with NARVAL could only be caused by a bright object, 
which would have been easily detected spectroscopically. Hence, the M giant itself must be 
magnetically active.

\subsection{Metallicity and atmospheric parameters determination}

We used our NARVAL spectrum of 13 March 2009 to derive the metallicity 
and other parameters of EK~Boo. For this purpose, we measured equivalent 
widths of the Fe\,{\sc i} and Ni\,{\sc i} lines in the ``window'' 
7400 - 7580~\AA, a spectral region relatively free of molecular absorption. 
Values for the respective oscillator strengths $\log gf$ were taken, 
if available, from Gurtovenko \& Kostyk (1989). Otherwise, we used the 
values derived by Smith \& Lambert (1985) or those offered by the VALD 
database (Kupka et al. 2000). 

We then used spherically symmetric, LTE, 
hydrostatic model atmospheres as computed with the MARCS code (Gustaffson 
et al. 2008). The CNO abundances in atmospheres of M giants are known to 
be modified through the first dredge-up (see, e.g., Charbonnel 1994, 
Boothroyd \& Sackmann 1999 and sect. 4): $^{12}$C is depleted and 
$^{14}$N  is enhanced. We therefore used MARCS models for moderately 
CN-cycled composition with [C/Fe]=$-0.13$, [N/Fe]=$+0.31$, and 
$^{12}$C/$^{13}$C\,=\,20.          
This chemical composition agrees with the abundance ratios 
measured in field M giants by Smith \& Lambert (1990): 
[C/Fe]=$-0.19\pm 0.13$ and [N/Fe]=$+0.28\pm 0.11$. 

The stellar mass was taken to be $M=2\,M_\odot$ (see Sect.~4). But as 
shown by Plez (1990) and Garc\'{\i}a-Hernandez (2007), the temperature 
and pressure structures of model atmospheres are practically identical for 
models of $1\,M_\odot$ and $10\,M_\odot$ stars. 
The current version of the LTE line analysis and spectrum synthesis code 
{\sc moog} (Sneden 1973) was used here.

The microturbulence velocity was determined from the Fe\,{\sc i} lines 
by requiring that the derived abundances shall be independent of the equivalent 
width. The effective temperature was derived by requiring that the abundance 
calculated for the Fe\,{\sc i} lines shall not show any dependence on 
excitation potential (Fig.~5). The atomic line list, measured equivalent 
widths, and derived abundances are listed in Table~2. 

In the way described above, we derive an effective temperature of  
$T_{\rm eff}=3400$~K and a microturbulent velocity of $\xi=2.0$~km\,s$^{-1}$.  
In adopting these values, a surface gravity of $\log g = 0.0$ 
(estimated from $M_{\rm bol}$, $T_{\rm eff}$  and the stellar masses 
obtained by comparison with evolutionary tracks, see Sect.~4 and Fig.~7), 
we then derived the following abundances: 
$\log\varepsilon({\rm Fe})=7.38\pm 0.13$, 
$\log\varepsilon({\rm Ni})=6.12\pm 0.18$ or 
[Fe/H]$=-0.14\pm 0.13$ and 
[Ni/H]$=-0.13\pm 0.18$\footnote{We adopt here 
the usual spectroscopic notation [X/Fe]$=\log(N_{\rm X}/N_{\rm Fe})_\star - 
\log(N_{\rm X}/N_{\rm Fe})_\odot$.} 
where the solar abundances of Fe and Ni are from Anders \& Grevesse (1989).
These values indicate a metallicity typical of M giants.

The accuracy of the above-derived metallicity of EK~Boo depends on 
the uncertainties in the adopted atmospheric parameters. Spectral signatures 
of singly ionized elements are highly sensitive to the variations in 
$\log g$, and to a lesser extent to the variations of $T_{\rm eff}$.
Small changes in the microturbulent velocity lead to larger changes in the 
derived abundances. Table~3 lists the abundance changes that are caused by 
individual changes in the various atmospheric parameters. -- In summary, 
we can state that the metallicity of EK~Boo is comparable to the solar one.

\subsection{Lithium abundance} 

In order to determine the lithium abundance of EK~Boo, we used the 
Li\,{\sc i} resonance doublet at 6707.8~\AA. In this region, the spectrum 
is heavily blanketed by TiO molecular lines, especially by the $\gamma$ 
electronic transition. In our computations of synthetic spectra we used 
the TiO line list given by Plez (1998), restricted to $^{48}$TiO. In deed,
when Plez et al. (1993) computed a restricted line list for the $\gamma$ 
system of TiO around the 6708~\AA\ Li\,{\sc i} line for all five isotopes 
of Ti from $^{46}$Ti to $^{50}$Ti, they found that even if they included 
all isotopes to improve the match between calculated and observed spectra, 
this did not change the derived Li abundance.
 
Vacuum values of wavelengths ($\lambda_{\rm vac}$) were transformed to
air wavelengths ($\lambda_{\rm air}$) according to the dispersion formula 
(see Birch \& Downs 1994, Morton 2000). For the dissociation energy of 
TiO, a value of $D_0=6.87$~eV was used. CN lines in the vicinity of the 
Li\,{\sc i} doublet were included in our line list, even though the CN red 
system lines near 6700~\AA\ are very weak in M giants (Luck \& Lambert 1982).
Wavelengths and oscillator strength for the individual hyperfine and 
isotopic components of the lithium lines were taken from Hobbs et al. (1999). 
A solar $^6$Li/$^7$Li isotopic ratio ($^6$Li/$^7$Li$=0.081$) was adopted in 
the computations of our synthetic spectra. Atomic line data were taken 
from the VALD database (Kupka et al. 2000).

As mentioned above, the TiO veiling effect in the EK~Boo spectrum is 
significant. This makes it difficult to place the continuum. For the local 
continuum, we adopted the highest point of the spectrum near the Li\,{\sc i} 
line, at a wavelength of 6710.7~\AA. Both the observed and synthetic spectra 
around 6708~\AA\ line are shown in Fig.~6. 

We cannot find any traceable presence of the Li\,{\sc i} 6708~\AA\ resonance 
line in the spectrum of EK~Boo. Using alternative synthetic spectra, an 
upper limit to the lithium abundance of $\log\varepsilon({\rm Li})\le -0.8$ 
was determined.

%

\begin{figure} 
   \centering
   \includegraphics[width=7.5cm]{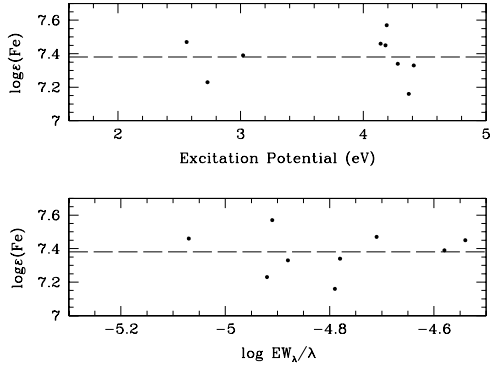}
   \caption{Iron abundance as derived from individual Fe\,{\sc i} lines,
    $\log \varepsilon$(Fe), {\it versus} excitation potential and reduced 
    equivalent width, $\log~(W_{\lambda}/\lambda$). An atmospheric model with
    $T_{\rm eff}$= 3\,400\,K, $\log g$ = 0.0, and $\xi$ = 2.0 km\,s$^{-1}$ 
    was used. }
   \label{Figure5}
\end{figure}

\begin{figure} 
   \centering
   \includegraphics[width=7.5cm]{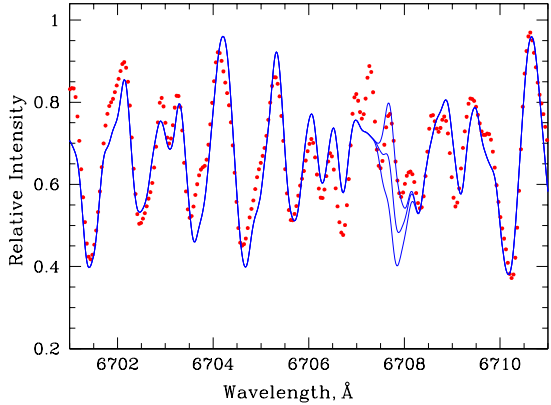}
   \caption{Observed (dots) and synthetic (solid lines) spectra for EK~Boo in 
    the region around the  Li\,{\sc i} 6708~\AA\ resonance line, the 
    observed spectrum is scaled in intensity for a good match. The synthetic 
    spectra represent lithium abundances of 
    $\log\varepsilon({\rm Li})= -1.0,$ 0.0, and +1.0. }
   \label{Figure6}
\end{figure}

\begin{table}[!htb]
\caption{Observed Fe\,{\sc i} and Ni\,{\sc i} lines, measured equivalent widths (EW),
         and derived abundances.}
\label{table:X1}
\begin{tabular}{cccccc}\hline\hline
Element &  $\lambda$\,(\AA) &  $\chi$(eV) & $\log gf$ & EW (m\AA) & $\log\varepsilon({\rm Fe})$\\
\hline
\ion{Fe}{I}& 7443.022  & 4.19 & $-1.810^1$ & $\;91.2$  & 7.57 \\ 
           & 7461.520  & 2.56 & $-3.570^1$ & 145.7     & 7.47 \\ 
           & 7498.530  & 4.14 & $-2.240^1$ & $\;63.3$  & 7.46 \\ 
           & 7507.266  & 4.41 & $-1.090^1$ & $\;99.0$  & 7.33 \\ 
           & 7511.019  & 4.18 & $\;0.099^2$ & 214.8    & 7.45 \\ 
           & 7531.144  & 4.37 & $-0.610^1$ & 122.1     & 7.16 \\ 
           & 7540.430  & 2.73 & $-3.870^1$ & $\;91.0$  & 7.23 \\ 
           & 7568.899  & 4.28 & $-0.890^1$ & 124.9     & 7.34 \\ 
           & 7583.787  & 3.02 & $-1.930^1$ & 199.3     & 7.39 \\ 
\ion{Ni}{I}& 7393.600  & 3.61 & $-0.100^1$ & 173.9     & 6.28 \\ 
           & 7414.500  & 1.99 & $-1.970^3$ & 220.5     & 6.06 \\ 
           & 7555.598  & 3.85 & $\;0.100^1$ & 137.8    & 5.90 \\ 
           & 7574.043  & 3.83 & $-0.470^1$ & 126.9     & 6.26 \\ 
\hline
\end{tabular}
\begin{list}{}{}{}{}{}{}
\item[$^1$] Gurtovenko \& Kostyk (1989);
\item[$^2$] VALD data base;
\item[$^3$] Smith\& Lambert (1985).
\end{list}
   \end{table}

\begin{table}  
\caption{Abundance uncertainties for EK~Boo. The second and third columns 
give the variation of the abundance that are caused by using a different 
$T_{\rm eff}$, i.e. larger or smaller by 100K. The other columns refer to changes in $\log g$ and $\xi$ 
respectively.}
\label{table:3}
\begin{tabular}{lccccccc}\hline\hline
Species & $\Delta$ $T_{\rm eff}$ & $\Delta T_{\rm eff}$ & $\Delta\log g$ & $\Delta\log g$ & $\Delta\xi$ &
$\Delta\xi$& \\
$_{\rule{0pt}{8pt}}$ & $+100$~K & -100~K & +0.5 & -0.5 & +0.3 & -0.3 \\
\hline             
Fe\,{\sc i}    & -0.12 & +0.12 & +0.18 & -0.20 & -0.16 & +0.18 \\
Ni\,{\sc i}    & -0.11 & +0.11 & +0.19 & -0.20 & -0.21 & +0.25 \\
\hline
\end{tabular}
\end{table}



\section{Mass and evolutionary status of EK~Boo}

In Sect.~3.3 we derive an effective temperature of 3400~K for EK~Boo. In the 
literature, values of 3420~K and 3577~K are given, according to Perrin et 
al. (1998) and Dyck et al. (1998), respectively. Hence, we here determine 
the mass and evolutionary status of EK~Boo on the basis of both these 
effective temperatures. We use the parallax from the New Reduction 
Hipparcos catalogue by van Leeuwen (2007), the $V$ magnitude  from the 
1997 Hipparcos catalogue, 
and bolometric corrections according to Buzzoni et al. (2010).

Figure~7 shows the positions of EK~Boo (black circles, for the two 
$T_{\rm eff}$ quoted above) and other M giants (see Table 4) in 
the HR diagram, in comparison to standard stellar evolutionary tracks.
These have been computed with the code STAREVOL by Lagarde \& 
Charbonnel (in preparation) for solar metallicity (with Asplund et 
al. 2005 chemical composition), ignoring stellar rotation. Tracks are 
shown for models with masses between 1.0 and 5.0~$M_{\odot}$. 

We focus here on the luminosity and effective temperature range of our sample 
stars, so that only the relevant RGB and AGB sequences (dashed and solid 
lines, respectively) appear on the graph. While low-mass stars 
($\leq$ 2.2~$M_{\odot}$) ignite He-burning in a 
degenerate core at the tip of the RGB, i.e., at very high luminosity and 
cool effective temperature, more massive stars start central-He burning 
at much lower luminosities and hotter $T_{\rm eff}$. Hence, the short 
RGB-tracks of the latter fall outside Fig. 7, showing only their AGBs. 
Respectively, for the low-mass stars we have plotted only the RGB sequences. 
As an exception, the 2.2~$M_{\odot}$ model is shown with both RGB and AGB.

When an effective temperature of EK~Boo of 3400 K is considered, this 
comparison leads to an estimate of the mass of 3.1$\pm$0.5~M$_{\odot}$. The uncertainty quoted here reflects only the parallax error.  
And in this case, EK~Boo is found to be evolving already on the AGB. At the luminosity of EK~Boo our 3~$M_{\odot}$ model has 
just undergone a first and relatively weak thermal pulse. By contrast, an 
alternative effective temperature of 3577 K would lead to a lower initial 
mass of 2.3$\pm$0.3~$M_{\odot}$, when compared to the tracks. Then the 
star would be either close to the tip of the first ascend (i.e., on the red 
giant branch, RGB), or on the AGB near the onset of the thermal pulse phase. 
In these two cases (i.e., 3.1 or 2.3~M$_{\odot}$, for  $T_{\rm eff}$=3400  
or 3577~K, respectively), the corresponding stellar radii are 200 and 109~R$_{\odot}$, respectively if one uses the 2007 parallax of EK~Boo. 

According to Auri\`ere et al.~(2009, see their Fig.~5), consideration of
stellar rotation would neither change the derived mass value, nor the 
evolutionary status for such a bright giant. Let us point out, however, that 
the models presented here were computed without core overshooting. For a 
given stellar mass, introducing core overshooting actually leads to a more 
massive He-core at the end of the main sequence, and thus to He-ignition 
at a somewhat lower RGB luminosity. In that case, the RGB tip-luminosities 
for stars above $\sim$ 1.5~$M_{\odot}$ could actually be lower than those 
shown in Fig.~7. This makes an even stronger case for EK~Boo being an AGB 
star. 

\begin{figure} 
   \centering
\includegraphics[width=8cm, angle=0]{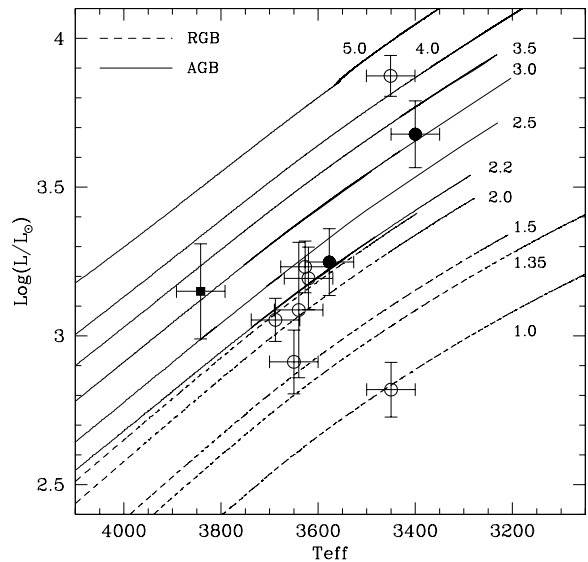}
      \caption{Position in the H-R diagram of EK~Boo and other M giants, 
adopting $T_{eff}$ from the literature, parallaxes from the New Reduction 
Hipparcos catalogue by van Leeuwen (2007), $V$ magnitudes  from the 1997 
Hipparcos catalogue, and bolometric corrections from Buzzoni et al. (2010).
EK~Boo's position (filled circles) is shown for two different effective 
temperatures (3420K and 3577K, see text). $\beta$ And, another Zeeman detected 
M giant, is represented by a filled square. Standard evolutionary tracks for 
solar composition and various initial stellar masses (as indicated) are also
plotted (Lagarde \& Charbonnel, in preparation). Dotted lines show the 
evolution up to the tip of the first ascent (the RGB) for the low-mass 
models (1.0 to 2.2~M$_{\odot}$), while AGBs are indicated by solid lines for the intermediate-mass stars (2.2~M$_{\odot}$ and above).
For clarity, only the 2.2~M$_{\odot}$ model is shown with both RGB 
and AGB.} 
         \label{Figure7}
   \end{figure}

Major uncertainties on the determination of the mass and evolutionary 
status of our other sample stars actually come from the errors on the 
effective temperature and on the parallax of the star, as can be seen in 
Fig.~\ref{Figure7}. However, uncertainties of the bolometric corrections 
are also significant. Here, we use the prescription by Buzzoni et al. (2010), 
who provide a fitting function for $BC_V$ vs. $T_{\rm eff}$, supposed to be 
valid in the range of $3300 \le T_{\rm eff}\le 5000$~K. For 
$T_{\rm eff}= 3400$~K, as derived above for EK~Boo, this gives $BC_V=-3.255$. 

Buzzoni et al. (2010) show that for stars with $T_{\rm eff} \ge 4000$~K  
the $BC_V - T_{\rm eff}$ relation does not depend on stellar 
metallicity as far as atomic transitions prevail as the main source of metal opacity in 
their spectra. However, for a star as cool as $T_{\rm eff}=3400$~K, 
molecular opacity (mainly due to TiO) becomes important in visual wavelength, 
which could in turn affect the derived $BC_V$ value. To the cool end ($T_{\rm eff} 
\le 3700$~K), the sample used by Buzzoni et al. (2010) is actually biased in 
this respect, since only giants from a metal-rich open cluster (NGC~6791, 
[Fe/H]=+0.4) were available for the bolometric correction determination.

Hence, we compared the Buzzoni-values of $BC_V$ with those which would follow 
for our M giants from Flower (1996) or Bessel, Castelli, \& Plez (1998). For 
less cool stars, i.e. $T_{\rm eff} \ge 3500$~K, all these bolometric 
corrections do not differ significantly from each other, but for cooler 
stars they do. For $T_{\rm eff}= 3400$~K, as derived for EK~Boo, 
we get a $BC_V=-2.70$ according to Flower (1996), and --3.07 when using 
Bessel et al. (1998). Taking into account these significant uncertainties 
in BC, the best we can conclude is that the mass of EK~Boo lies in the 
range of 2.0--3.6~$M_\odot$.

The absence of the $^{7}$Li enrichment in the atmosphere of EK~Boo is 
consistent with what is expected for a star within this mass range and at 
this evolutionary stage. The star has indeed depleted in the previous phases 
of its evolution all the $^{7}$Li it was born with, and it is not massive 
nor bright enough to produce fresh $^{7}$Li via the so-called Cameron \& 
Fowler (1971) mechanism on the TP-AGB (see e.g. Smith 2010 and reference 
therein)
\footnote{Sackmann \& Boothroyd (1992)  predict that 
Li-rich AGB stars will have a mass range of 5 to $7\,M_\odot$
for a solar metallicity.}.

\section{Rotation and dynamo operation in EK Boo}

\subsection{Rotation}
As for most M giants, the rotational period of EK~Boo is not known. 
But in this case, $v\sin i$ has been determined. The CORAVEL 
value obtained on the basis of the cross-correllation method (Melo et al. 2001) is $11\pm 1$~km\,s$^{-1}$ (H\"unsch et al. 2004). 
Our NARVAL data and the spectral synthesis method (permitting us to account
for rotational and macroturbulent broadening) yield $v\sin i$ of
8.5~km\,s$^{-1}$ with an accuracy 0.5~km\,s$^{-1}$.
To estimate the projected rotational velocity, several synthetic spectra
were generated by {\sc moog} with different values of $v\sin i$ and
$v_{\rm macro}$.
The best fit was found for the following combination of the line-broadening factors: $v\sin i$ = 8.5~km\,s$^{-1}$ and $v_{\rm macro}=2.0$~km\,s$^{-1}$. An adoption of low projected rotation velocity combined with high macroturbulent velocity leads to synthetic spectra that are noticeably
different from the observed one. However, we have to mention that
we do not know the real geometry of macroturbulent motions in the
atmosphere of EK~Boo. In our calculations we used an assumption of radial-tangential macroturbulence profile (Gray 1992). 

The 
difference between these two values could simply arise because of the different 
methods of $v\sin i$ determination. In general, M giants as very evolved 
stars are expected to be slow rotators. In this respect, both values above 
are surprisingly high! Hence, EK~Boo appears to be a single M giant 
with a relatively fast rotation, compared to the statistical value for its 
spectral class (Zamanov et al. 2008).

An estimate of the rotational period $P$ of the star (upper limit only) could 
be made using the radius of the star, as derived from its angular diameter. 
Adopting $\theta$ of 7.90 $\pm$ 0.40 mas (Dyck et al. 1998), as well as a 
distance of 248 $\pm$ 25 pc (by the new Hipparcos parallaxes of van Leeuwen, 
2007), we obtain a radius of 210 $\pm$ 21 $R_\odot$. This value is about 
twice as high as the statistical value for an M5 giant (120 $R_\odot$) 
given in Dumm \& Schild (1998). It is however consistent with the value of 
200 $R_\odot$ derived in Sect.~4 for EK~Boo being an AGB star with 
a $T_{\rm eff}$ of 3400~K. Assuming rigid body rotation and vsini of 8.5 km/s, 
we obtain $P \le 1247^d$ for the empirical radius of 210 $R_\odot$, and 
$P \le 714^d$ for the statistical radius of 120 $R_\odot$. 
For vsini = 11 km/s, these values are $964^d$ and $552^d$, respectively.

Of course, any tilt of the giant's axis into the line of sight would 
in deed mean it is rotating even faster than suggested by the radial 
velocity measurement. Adopting the statistical value of $\sin i = \pi/4$
and the empirical radius, we then obtain a period  of $846^d$ for 
$v\sin i = 8.5$~km\,s$^{-1}$ (or $654^d$ for $v\sin i = 11$~km\,s$^{-1}$).

In future, our observations and Zeeman Doppler Imaging (ZDI) method will 
be able to shed more light on the actual rotational period of EK~Boo.

\subsection{Dynamo operation and its origin}

Indirect evidence for the magnetic activity of EK~Boo has been reported before our 
NARVAL study. The star attracted the attention of H\"unsch et al. (1998) 
because of its high X-ray luminosity. Later, H\"unsch et al. (2004) studied its 
variability in X-ray and optical activity indicators. They too ruled out 
the possibility of a symbiotic nature of EK~Boo.

Our study (see Sect.~3.1) revealed a variable longitudinal magnetic field in EK Boo for which a dynamo could be a possible explanation. A dipole (as a remnant of an Ap star on the main sequence) is an unlikely explanation, considering the big radius and big convective envelope at this evolutionary stage. We now attend to the question about the possible origin of the dynamo and its type in this evolved star.

 
A star of about 2.2--3.6~$M_\odot$ at the evolutionary stage of EK~Boo has 
just entered the thermally-pulsing phase on the AGB. It has a complex internal 
structure with an O/C core surrounded by He- and H-burning shells, and a very 
deep and extended convective envelope.

Turbulent motions and fast rotation could result in dynamo 
action of the $\alpha- \omega$ type, as described in Nordhaus et al. (2008).
Could such a dynamo explain the magnetic field in EK~Boo? We calculated the 
Rossby number (R$_o$), which is indicative for an efficient dynamo action 
of this type. For the statistical rotational period of 846~d (see Sect. 5.1) 
and a convective turnover time $\tau_c$ of 182~d, as obtained from the model of 
2.5~$M_\odot$ described in Sect.4, we determined a R$_o$ of about 4.7. 
For a period of 654~d (as for a smaller, statistical radius, see Sect. 5.1),
the value of R$_o$ would be 3.6. For these values, we cannot expect a very 
efficient $\alpha - \omega$ dynamo for EK~Boo, but this dynamo might be 
still operationable (see Soker \& Tylenda 2007). Our future observations and 
a direct determination of the rotational period could lead to a better 
knowledge of the Rossby number and the $\alpha - \omega$ dynamo efficiency 
in this M giant. 

Considering the high density and height difference between the 
photospheric layers of the giant and its bottom of the convection zone deep 
inside, it is well possible that the deep layers rotate a lot faster then 
the surface, which suffers from magnetic braking. In this way, 
the $\alpha - \omega$ dynamo may yet be more efficient than 
expected on the basis of the surface rotational rate.
 
The question remains of where the required angular momentum should 
come from to run this dynamo. Schr\"oder et al. (1998) studied the 
evolutionary status of evolved stars that have coronal X-ray emission, which is a 
good indicator of stellar activity among less cool giants. And 
Konstantinova-Antova et al. (2009) studied fast rotating single G and K giants 
with detected magnetic field. The most active giants in their samples are 
above 1.5~$M_{\odot}$, at different evolutionary stages after the main 
sequence (MS): first gap crossers, stars near the base of the RGB and its 
lower part. These stars had no convective outer layers and, consequently, no 
stellar activity on the MS. Hence, magnetic braking cannot yet have 
consumed  their fast rotation and fresh activity has set in at these 
early post-MS stages. By contrast, EK~Boo has evolved much further and 
its convective envelope is a lot more extended. It should have slowed 
down long ago, and fresh angular momentum is required here to run its faster rotation 
and a dynamo. 

The possible source is the contracting core region of this giant: Without 
any angular momentum transfer, the contracting core would be considerably speeding up 
its rotation rate, both in an RGB and an AGB giant of such 
a luminosity such as EK~Boo. Its final product, a white dwarf, would have 
to rotate as fast as in minutes, which is physically impossible. Instead, 
we now have even direct evidence (see Charpinet et al. 2009) that white 
dwarfs are rotating a lot slower, with periods of the order of a day. 
Hence, evidently over 99\% of the angular momentum of the core region is 
lost during the late evolutionary stages. As discussed by Talon \& 
Charbonnel (2008), internal gravity waves may strongly contribute to 
the angular momentum redistribution in intermediate-mass stars such as 
EK~Boo, when they reach the early-AGB phase. Hence, it may serve the 
expanded convective envelope to keep it spinning at a sufficient 
rate to run a dynamo process in M giants.

The other possible dynamo type operating in AGB giants is the turbulent 
dynamo, where the turbulent motions play the main role in the field 
amplification process and in the restoration of the poloidal component 
(Brandenburg, 2001). However, for EK~Boo this dynamo is less 
efficient than the $\alpha - \omega$ one. Soker \& Zoabi (2002) estimated 
that in AGB stars the magnetic field can reach only values of less than 1~G. 
This is inconsistent with the much larger net field strength, which we 
observe in EK~Boo.

A second explanation for the magnetic field in this M giant could be 
given if EK~Boo were a merger of a binary system, which would now be in its spin-down 
phase. Then an $\alpha - \omega$ dynamo is well possible to 
operate (Soker \& Tylenda 2007). The large radius of EK~Boo together with 
the high $v\sin i$ appears to support this idea. But for the moment, the 
accuracy of the angular diameter measurement and of the parallax are not 
high enough to give a reliable determination of the radius of the star. 
In future, we will need more precise values for these parameters, to be able 
to distinguish between these two possibilities for $\alpha - \omega$ dynamo 
operation in the star.



Another possibility for spinning up a giant and to induce dynamo action is 
planet engulfment (Siess \& Livio, 1999).  In this case, according 
to Melo et al. (2005), there should be an enhanced Li abundance.
Indeed, Drake et al. (2002) found that among the rapid rotating 
($v\sin i \ge 8$~km\,s$^{-1}$) single K giants, there is a very large 
proportion ($\sim$50\%) of Li-rich objects. By contrast, there is only
a very small fraction ($\sim$2\%) of Li-rich stars among the common type of 
slowly rotating K giants.

Unfortunately, we do not observe any enhanced lithium abundance
in EK~Boo, although we analysed this point carefully (see Sect.~3.4). 
However, note that only a recently engulfed big planet would 
cause a significant Li content on the star, since Li is a fragile element. The 
spin-down time, on the other hand, remains unclear, but it might exceed the
lithium life-time. Hence, we cannot completely exclude the possibility that 
the enhanced rotation might be a result of a big planet engulfment.

\section{Are there other magnetically active M giants?}

If dynamo operation on the AGB is a result of the evolution of the star 
rather than an exotic case like binary merging or planet engulfment, we 
should expect more giants to have a magnetic field at the stage of EK~Boo. 
As mentioned already in Sect.~1, there is indirect evidence for such 
activity, which stimulated us to begin observing several other M giants
of interest in that respect.

Hence, we observed eight other M giants with the same procedures as for 
EK~Boo (see Sect. 2). Three of them are known for their high X-ray 
emission: 15~Tri, 42~Her, and  HD~187372 = HR~7547 (H\"unsch et al. 1998, 
2004). The sources 15~Tri and HD~187372 are also known as spectroscopic binaries of
a long period (i.e., no synchronization), and 42~Her is considered to have 
a stable $RV$ (H\"unsch et al. 2004, Famaey et al. 2009). Our own NARVAL  
$RV$ measurements confirm these findings (the specific case of EK~Boo 
is discussed in Sect.~3.2). Five other M giants are fast rotators (Zamanov 
et al. 2008). These are HD~167006, HD~184786, 8~And = HD~219734, HD~18191, 
and $\beta$~And = HD~6860. All these giants show no radial velocity 
variations and may be considered to be single stars (Famaey et al. 2005). 

Data of all observed giants are presented in  Table~4. The effective 
temperatures are from the VizieR catalogues. The X-ray luminosity $L_x$ is 
from  H\"unsch et al. (2004); the measurements are both from the ROSAT and 
Einstein satellites. Positions in the HR diagram are shown in Fig.~7, 
assuming parallaxes from the New Reduction Hipparcos catalogue by van 
Leeuwen (2007), $V$ magnitudes  from the 1997 Hipparcos catalogue, and 
bolometric corrections from Buzzoni et al.~(2010).

All these are stars in the mass interval 1.0-4.5~M$_\odot$, situated near
the tip of the RGB or on the AGB. The least cool member of the 
sample, $\beta$~And is among the more massive stars in our study 
(3.1$\pm$0.5~$M_{\odot}$), and is located on the AGB. Note that at 
the adopted $T_{\rm eff}$ of $\beta$~And, the uncertainty on the $BC_V$ is 
much smaller than for EK~Boo. Consequently, the uncertainty of
its mass is mainly owing to the error of the Hipparcos parallax.

\begin{table*}
\caption{Data for the observed M giants.}             
\label{table:2}      
\begin{tabular}{c c c c c c c c c c c}        
\hline\hline                 
Star    &Other Name  &Sp class& T$_{eff}$ &$v\sin i$     & log$L_x$&Date          &$N^o$. exp.&Detection& $B_l$ & $\sigma$\\
        &            &        & K         & km s$^{-1}$ &         &              &       &           & G   & G       \\
\hline 
HD130144&EK~Boo      &M5III   & 3400      & 8.5/11      &30.30 - 31.15& See Table 1&       &    DD       &     &         \\
\hline
HD6860  &$\beta$ And &M0III   & 3842      & 5.6         &       &16 + 26 Sep.08&14     &MD         &-0.95& 0.16    \\
        &           &         &            &            &       &24 Sep. 09    &16     &nd         &-0.29& 0.10    \\
HD16058 &15 Tri     &M3III    & 3640       & 5.4         &30.8   &20 + 21 Sep.08& 5     &nd         &-0.68& 0.38    \\
        &           &         &            &             &       &20 Dec. 08    & 3     &nd         &-1.05& 0.59    \\
HD18191 &RZ Ari     &M6III    & 3450       & 9.6         &       &16 + 21 Sep.08& 5     &nd         &-0.89& 0.45    \\
HD150450&42 Her     &M2.5III  & 3650       & 2.5         &29.41  &19 + 30 Sep.08& 8     &nd         &-0.37& 0.19    \\
        &           &         &            &             &       &25 Feb.09     & 4     &nd         &-0.01& 0.31    \\
HD167006&V669 Her   &M3III    & 3688       & 5.2         &       &16 + 21 Sep.08&12     &nd         &-0.85& 0.32    \\
        &           &         &            &             &       &21 May 09     &16     &nd         &0.97 & 0.39    \\ 
HD184786&V1743 Cyg  &M4.5III    & 3451       & 7.8         &       &15 + 25 Sep.08& 8     &nd         &-0.16& 0.34    \\
HD187372&HR7547     &M1III    & 3620       & 4.4         &30.64  &19 + 29 Sep.08& 8     &nd         &0.31 & 0.34    \\
        &           &         &            &             &       &25 Feb.09     & 4     &nd         &-0.24& 0.47    \\
        &           &         &            &             &       &24 Nov.09     & 3     &nd         &-0.82& 0.53    \\
HD219734&8 And      &M2III    & 3627       & 4.9         &       &15 + 30 Sep.08& 8     &nd         &-0.28& 0.30    \\
\hline  
        &           &         &            &       &              &       &           &               \\
\end{tabular}

Individual columns list HD number, name of the star, spectral class, 
effective temperature $T_{\rm eff}$ from VizieR catalogues, 
$v\sin i$, log of  X-ray luminosity (in (erg\,s$^{-1}$), detection 
(DD = definite, MD = marginal, nd = no), $B_l$ and its error in G. 

\end{table*}

With the exeption of  $\beta$~And, which is much brighter, these M giants 
are in the same range of magnitude as EK~Boo. For these reasons, their magnetic field detection and $B_l$ measurements 
are of a similar accuracy as those of EK~Boo. 
The three fast rotators (HD18191, HD184786, HD219734)
were observed only in one month (September 2008). The sources $\beta$~And and HD167006 were observed during two months.  
The three X-ray luminous giants were observed during two or three months: September and 
December 2008 (15~Tri), September 2008 and February 2009 (42~Her), and 
HD~187372 was observed in September 2008, February 2009, and November 2009. 

A magnetic field was not detected with certainty for any of 
these stars within the limit of our accuracy (0.5~G). However, 
for $\beta$~And, the brightest star in this sample, we obtained a marginal 
detection of magnetic field (see Fig.~8) in September 2008, after averaging 
14 spectra. The measured $B_l$ is $-0.95 \pm 0.2$~G. New observations were 
obtained on 24 September 2009: a weak Stokes $V$ signal is still visible, 
but at a non significant LSD statistical level (Donati et al. 1997). 

This means, that magnetic fields may well be present in these stars, but with a 
weaker averaged longitudinal field, wherefore a different observational strategy 
might be necessary. However, a weak field seems an unlikely explanation for 15~Tri and HD~187372, which are X-ray sources as luminous as 
EK~Boo, but which have to date not yet yielded any Zeeman detection of a mean 
magnetic field. As suggested in Sect.~3.1, magnetic areas 
with opposite polarities on the surfaces of these stars could produce averaged 
longitudinal magnetic fields close to zero, and that possibly for extended 
periods. Actually, EK~Boo displays big differences in its magnetic field 
behaviour over an extended time interval (e.g. April -- December 2008). Hence, 
we will need to monitor all suspected M giant stars both deeper and for
longer periods of time. 

  \begin{figure} 
   \centering
   \includegraphics[width=8.95cm]{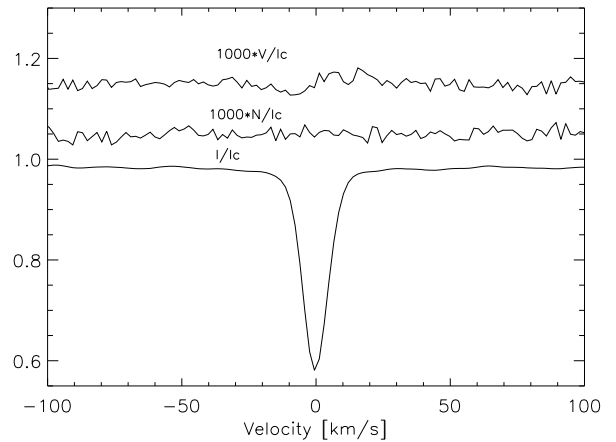}
   \caption{Averaged Stokes $I$ (bottom), $N$ (middle) and $V$ (upper) profiles for $\beta$~And. The spectra were obtained in September 2008.}
         \label{Figure8}
   \end{figure}

For the X-ray luminous M giants, it may not entirely be excluded that a
secondary star is the origin the X-ray radiation. Two of them, 15 Tri and HD187372 are 
long-period spectroscopic binaries (H\"unsch et al. 2004, Famaey et al. 
2009), and 42~Her is a visual binary. Their $v\sin i$, however, are smaller 
than the value of EK~Boo, which could simply mean that they are less active. 
All these stars have Ca\,{\sc ii} emission, but their K emission cores show 
only little variability, much less than what we observe in EK~Boo. 
This may support the idea that there is a link between their lower 
rotation velocity and a lower activity level in these three M giants. 

The five fast rotating single M giants also exhibit Ca\,{\sc ii} emission 
cores with a different strength. But we still lack sufficient observational 
data of them to judge their variability. In any case, all M~giants 
(with one exception: HD~18191) have lower $v\sin i$ values than EK~Boo and a lower activity level could not be excluded, too.

More observations will answer the question of magnetic field in these 
M giants. In the future, we also want to expand our sample of M giants 
with some Mira variables, for which magnetic field detection has been reported 
from maser emission lines (Herpin et al. 2006). In this way, we will be able 
to draw a more complete picture of magnetic activity in AGB stars, and we will 
finally know, whether EK~Boo is a special case, or rather whether this 
M giant is just the ``tip of an iceberg'' of magnetically 
active AGB stars.

\section{Conclusions}

   \begin{enumerate}
      \item The M5 giant EK~Boo was observed with NARVAL in the period 
of April 2008 to March 2009 during 10 nights. We analysed the $RV$ behaviour 
of the star and came to the conclusion that EK~Boo is likely a single star,
and that the observed variations of its $RV$ are due to some kind of 
pulsation. Its position on the H--R diagram reveals that EK~Boo is of 
2.0--3.6 $M_{\odot}$, located on the AGB or near the tip of the RGB. 
 \item
Magnetic field was definitely detected in the photosphere of EK~Boo and 
its longitudinal component $B_l$ was of the order of few Gauss. It is of 
a variable strength, and the activity indicators Ca\,{\sc ii} K\&H also 
show variability. EK~Boo is the first M giant for which surface 
magnetic field is detected.
      \item In March 2009, a complex structure in Stokes $V$ was observed.
This find may be indicative of a dynamo operation. In that case, EK~Boo could 
be the first M giant with proof of dynamo operation. 
           \item A dynamo of the $\alpha$ -- $\omega$ type is more likely 
to operate there than a turbulent dynamo. The driving mechanism for the 
$\alpha$ -- $\omega$ dynamo operation could be angular momentum dredge-up 
from the interior, binary merging, or perhaps a planet engulfment, but the 
absence of a high Li abundance in EK~Boo speaks against the latter (unless
the spin-down time is significantly longer than the Li destruction time). 
A turbulent dynamo appears to be challenged by the high measured field 
strength of EK~Boo.
       \item For one more M giant, $\beta$~And, we got a marginal detection 
of the magnetic field with a $B_l$ strength of --0.95~G. A different 
observational strategy is required to be able to detect a weaker magnetic 
field in our other sample M giants, and to answer the question, what kind of a special case EK~Boo is.

   \end{enumerate}

\begin{acknowledgements}
      We thank the TBL team for their service observing support. Our
observation time was granted under an OPTICON programme. We further acknowledge 
use of the Hipparcos database, the VizieR catalogue access tool (CDS, 
Strasbourg, France) and of the VALD data base in Vienna. R.K.-A. and I.S. 
acknowledge partial financial support under Bulgarian NSF contract 
DO 02-85/2009. R.K.-A. appreciates the possibility to work in LATT -- 
Tarbes as an invited researcher in the spring 2009. C.C. acknowledges 
financial support from the French Programme National de Physique Stellaire 
(PNPS) from CNRS/INSU and from the Swiss National Science Foundation (FNS).
\end{acknowledgements}

\end{document}